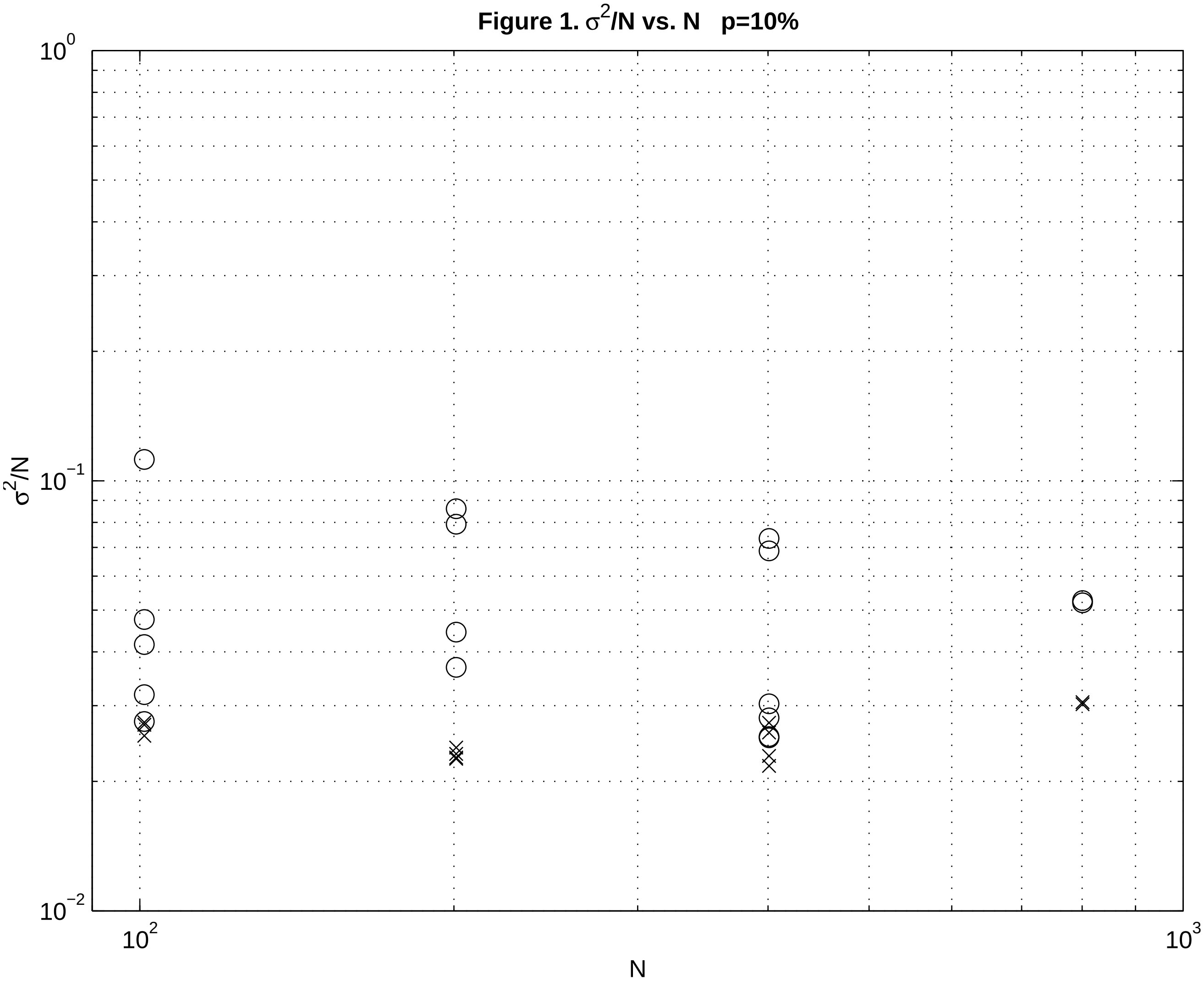

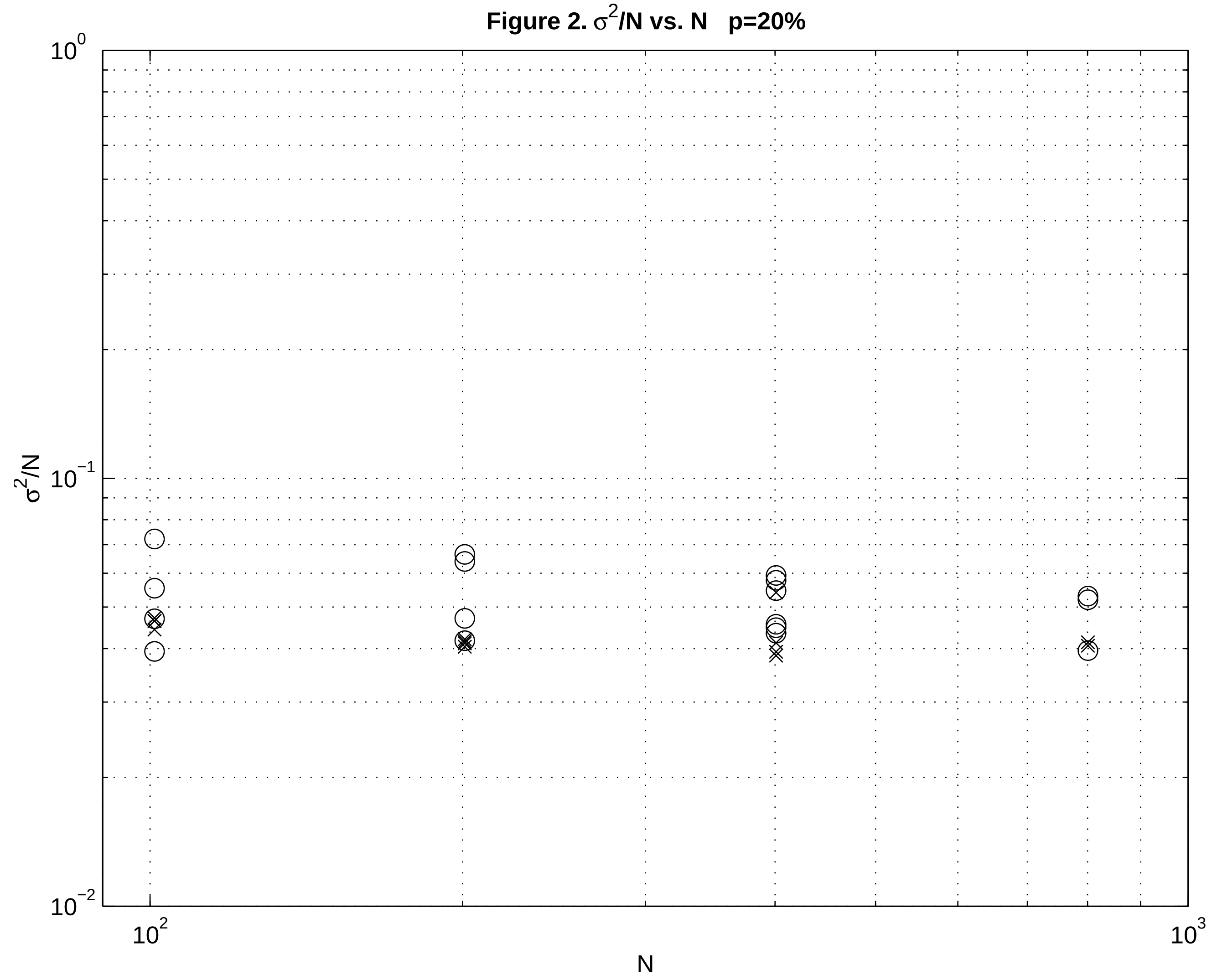

Figure 3. $\sigma^2/N$ vs. N p=40%

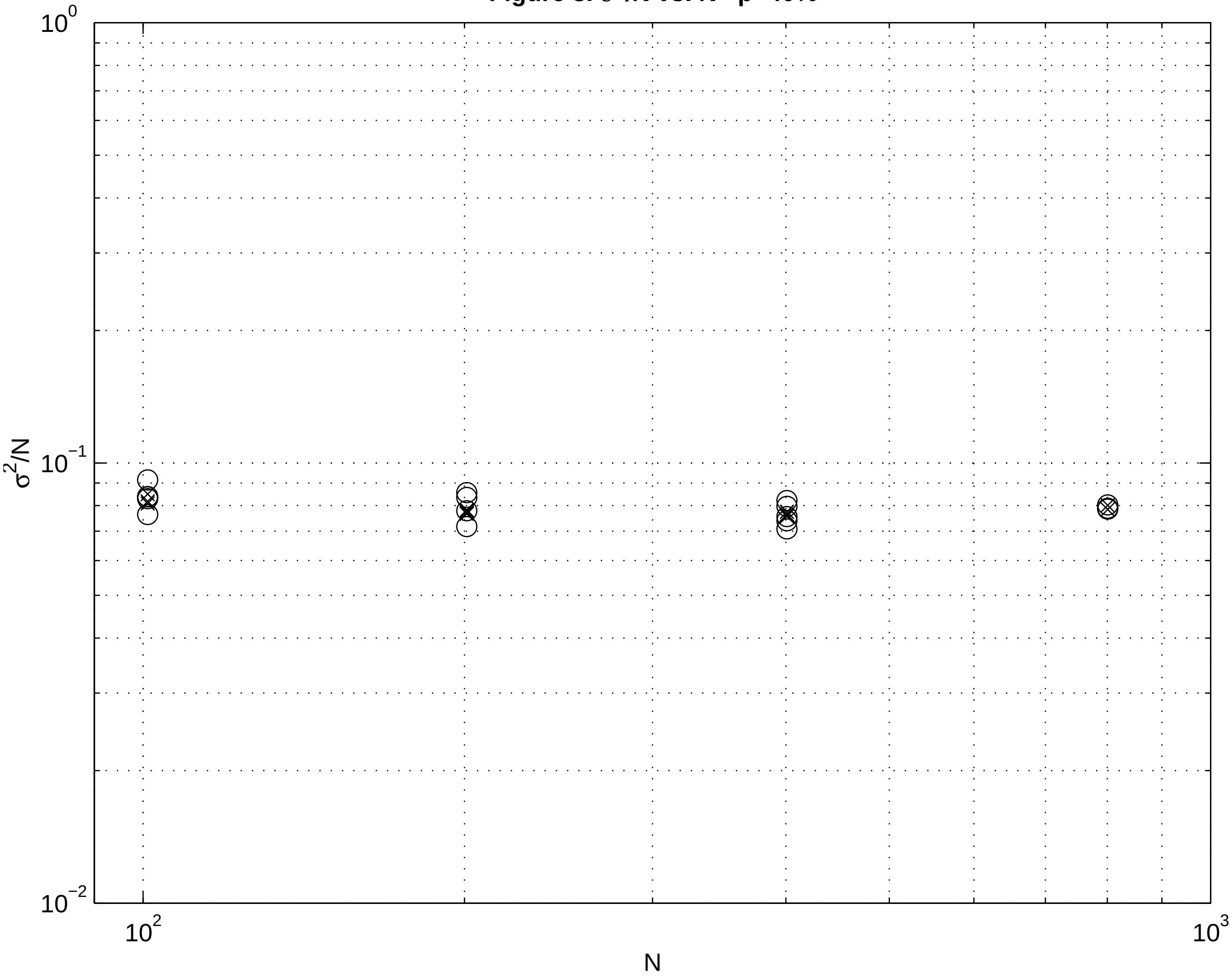

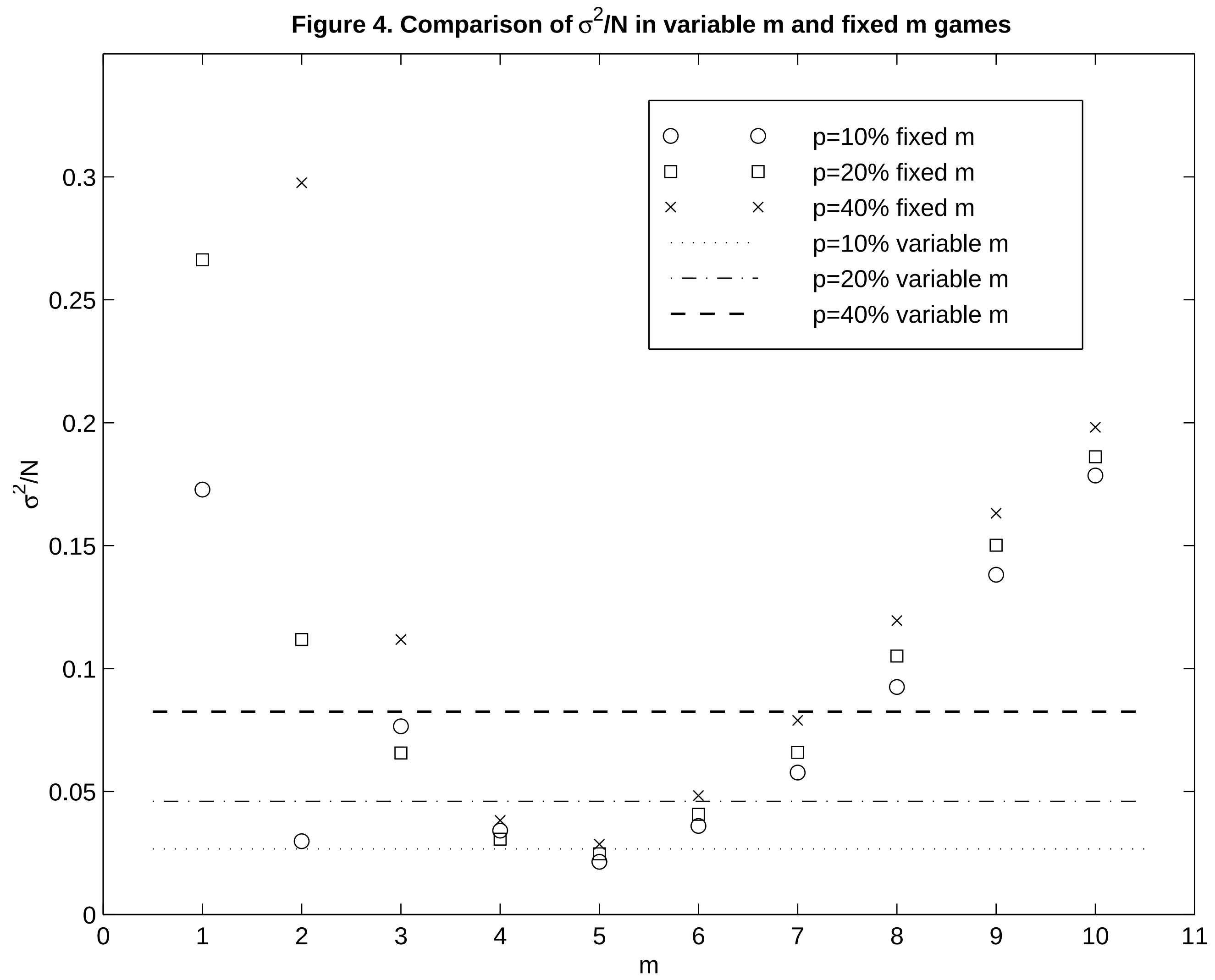
Figure 4. Comparison of σ2/N in variable m and fixed m games
p=10% fixed m
p=20% fixed m
p=40% fixed m
p=10% variable m
p=20% variable m
p=40% variable m
σ2/N
m
0.3
0.25
0.2
0.15
0.1
0.05
0
0 1 2 3 4 5 6 7 8 9 10 11

**Figure 5. $n_m$ and $w_m$ vs. m for a run with variant A dynamics**

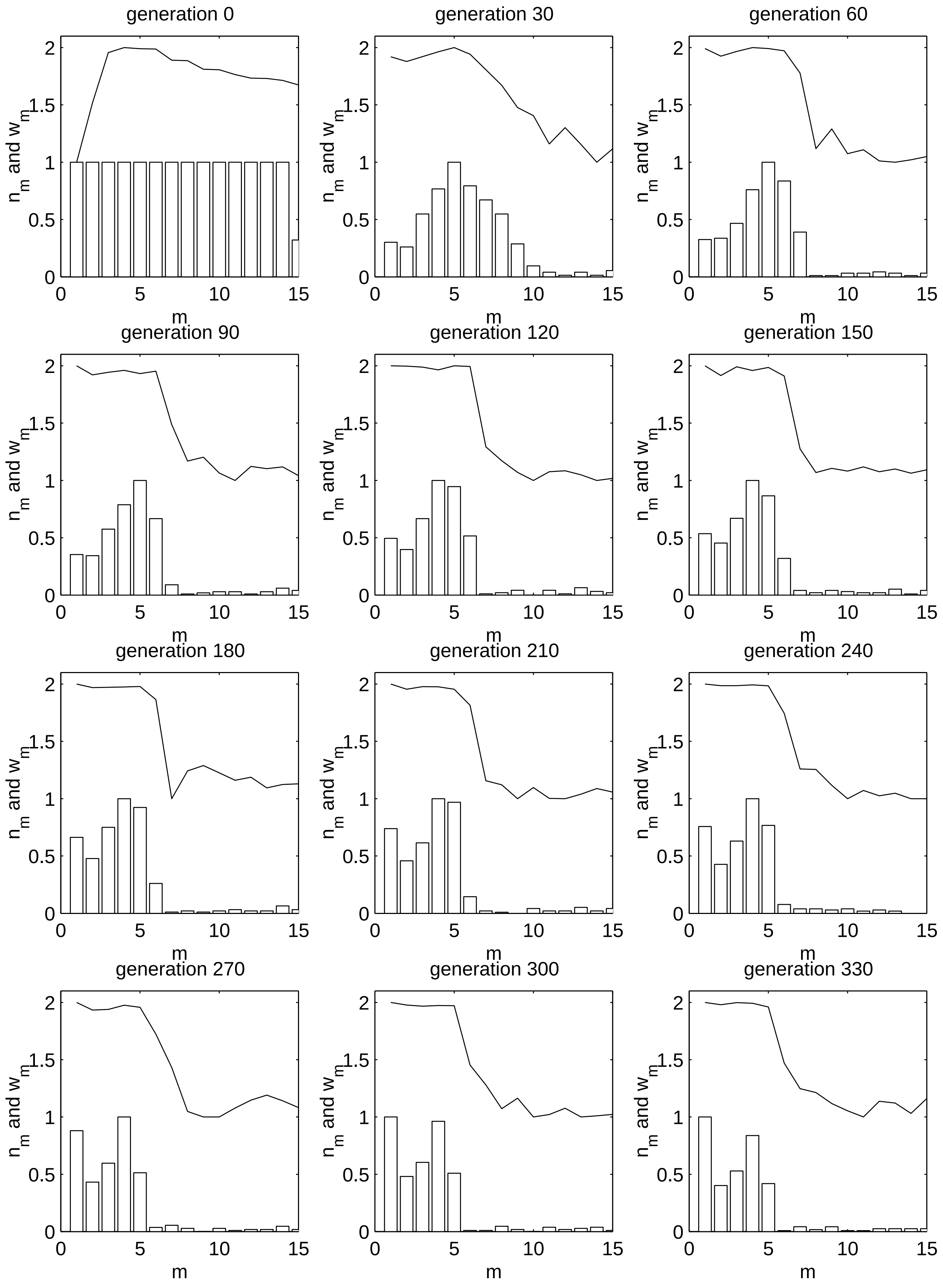

## Figure 6. $n_m$ and $w_m$ vs. m for a run with variant B dynamics

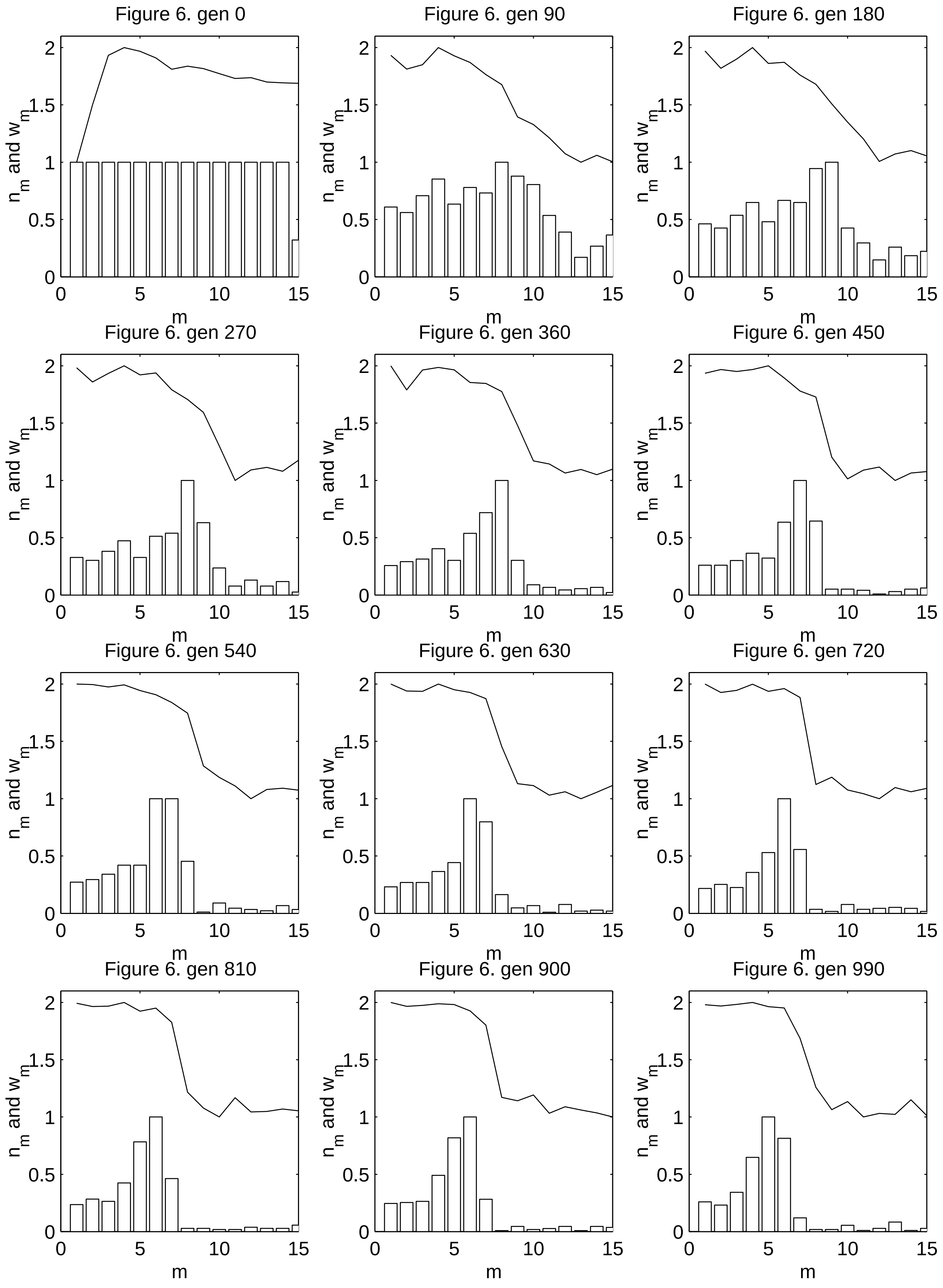

Figure 7. operationally defined $m_t$ vs. N

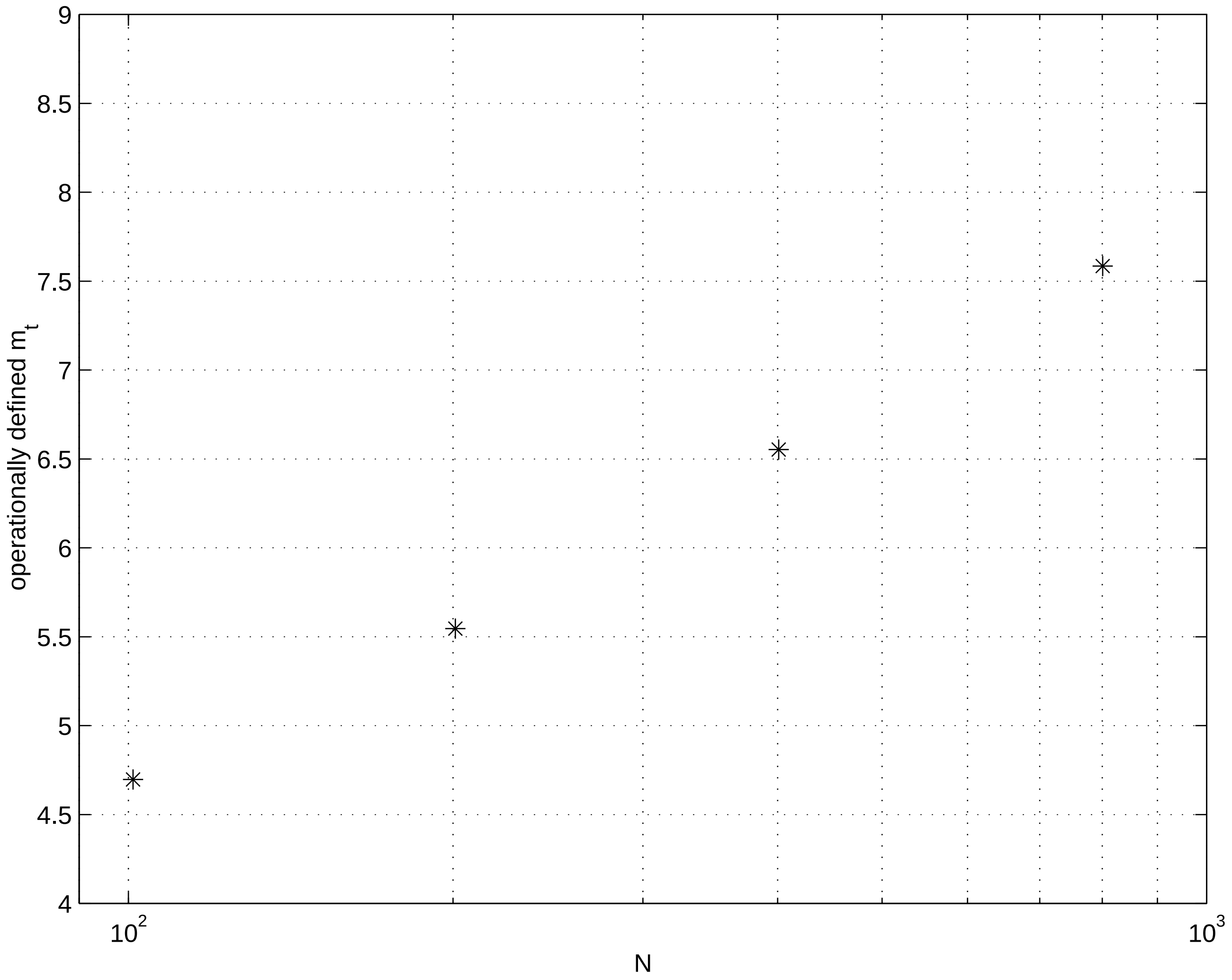

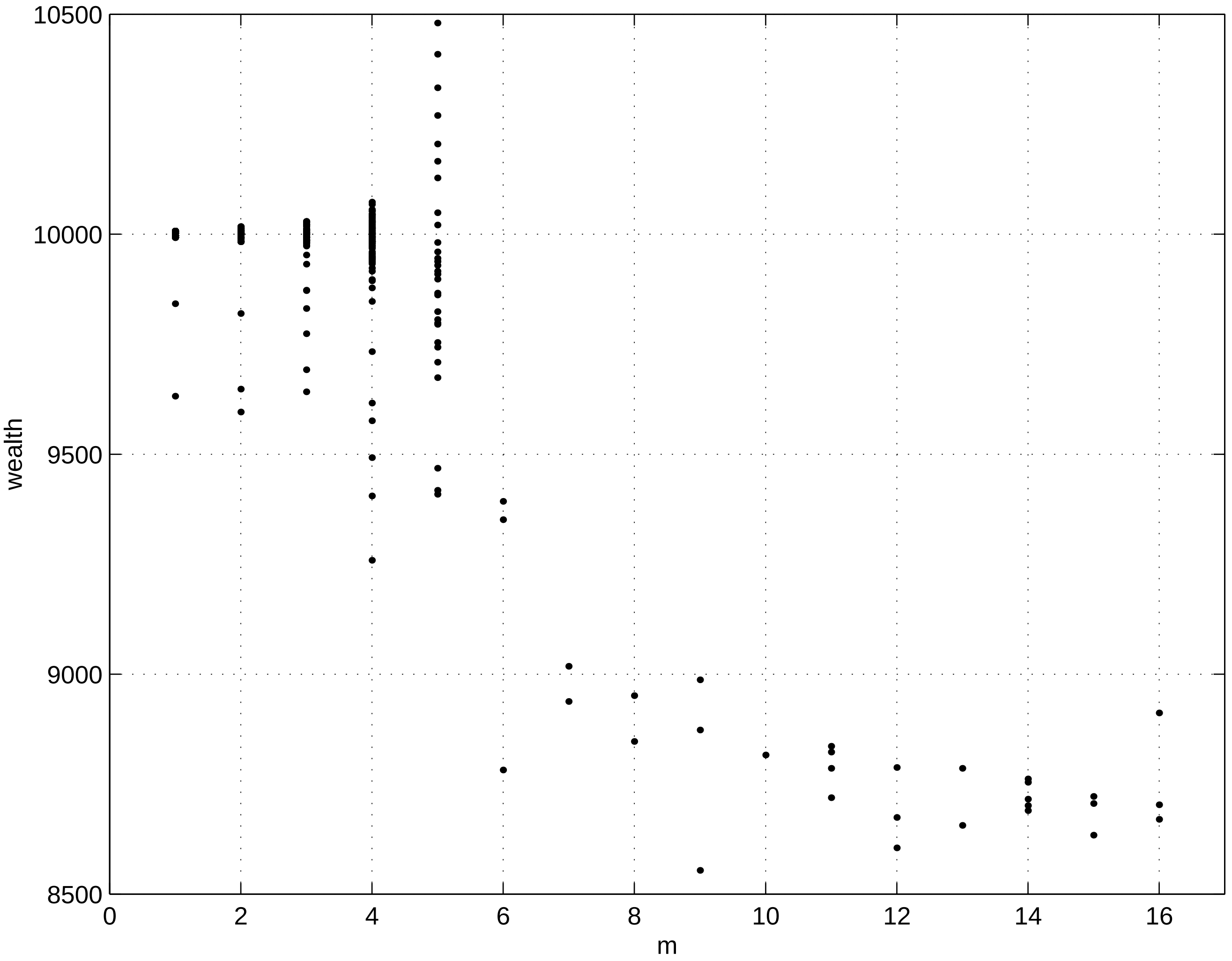

**Figure 8. typical example of wealth distribution vs. m**

# Evolution in Minority Games
# II. Games with Variable Strategy Spaces

Yi Li, Rick Riolo and Robert Savit
Program for the Study of Complex Systems and Physics Department
University of Michigan, Ann Arbor, MI 48109

Abstract

We continue our study of evolution in minority games by examining games in which agents with poorly performing strategies can trade in their strategies for new ones from a different strategy space. In the context of the games discussed in this paper, this means allowing for strategies that use information from different numbers of time lags, m. We find, in all the games we study, that after evolution, wealth per agent is high for agents with strategies drawn from small strategy spaces (small m), and low for agents with strategies drawn from large strategy spaces (large m). In the game played with N agents, wealth per agent as a function of m is very nearly a step function. The transition is at $m=m_t$, where $m_t \approx m_c-1$. Here $m_c$ is the critical value of m at which N agents playing the game with a fixed strategy space (fixed m) have the best emergent coordination and the best utilization of resources. We also find that overall system-wide utilization of resources is independent of N. Furthermore, although overall system-wide utilization of resources after evolution varies somewhat depending on some other aspects of the evolutionary dynamics, in the best cases, utilization of resources is on the order of the best results achieved in evolutionary games with fixed strategy spaces. Simple explanations are presented for some of our main results.

## I. Introduction

The problem of competition for scarce resources lies at the heart of many systems in the social and biological sciences. It is often the case that success in such a competition requires an agent adopting strategies that make his actions distinct from those of other players. Thus, in trade, one wants to be a seller when most agents are buyers (so that the price will be high) or a buyer when most agents are sellers (so that the price will be low). Compounding this drive to be different is the observation that in many systems agents are heterogeneous and generally adopt different strategies in their attempt to be different. Furthermore, the actions of the agents affect their environment, so that future choices of an agent in a heterogeneous population are conditioned by the past actions of the other agents, and by that agent's past experience in the context of the choices made by the collective.

This general structure was encapsulated in a problem posed by Arthur[1] which centered around the problem of attending a popular bar on nights when Irish folk music was being played at the bar. Many people want to go to the bar on such nights, but no one wants to go if there are more than a certain number of people at the bar, since then the place would be too noisy and no one would be able to enjoy the Irish music.

Motivated by Arthur's phrasing of the question, Challet and Zhang[2] suggested a simple model which has come to be known as the minority game, and which incorporates much of the basic structure of the kind of problem posed by Arthur. Versions of this model have been studied,[3,4,5,6,7.8.9,10]and much of the basic structure has been explicated, if not deeply understood. In particular, and most remarkably, it has been established that in games that are adaptive (but not evolutionary so that agents' strategies are fixed for the duration of the game), and in which agents can choose to exercise different strategies at different moments of the game, there can be an emergent coordination among agents' choices that leads to an optimum utilization of resources.[3,4] The controlling parameter in these games, z, is the ratio of the dimension of the strategy space from which the agents draw their strategies, $2^m$, to the number of agents playing the game, N. In these games, agents' strategies can use the publicly available information about which was the minority group for the previous m time steps in order to make their predictions of what will be the next minority group. If z is of order one, then there is good emergent coordination, and a good utilization of resources. If z is too small the agents' actions become maladaptive leading to a very poor utilization of resources, and if z is too large, overall resource utilization declines and approaches that of a collection of agents all of whom make random decisions.

Given the remarkable structure of the adaptive, non-evolutionary game, it is natural to ask what happens when evolutionary dynamics is included and agents are allowed to change their strategies under selective pressure. In examining the role of evolution, it is important to distinguish between two different cases. In the first, the strategy space available to the agents is fixed, so that all strategies of all the agents have the same value of m. Poorly performing agents can replace their strategies, but the new strategies must always be drawn from the same strategy space, and therefore have the same value of m. In the second case, the strategy space is allowed to vary. Different agents may possess strategies drawn from different strategy spaces (i.e. associated with different values of m). Moreover, poorly performing agents may replace their strategies with a different value of m.

The first case has been studied in a companion paper[6]. In that work we found that evolution typically improves system-wide utilization of resources. At the same time, many of the most intriguing general features of the adaptive, non-evolutionary games persist when evolution is incorporated. In particular, the best overall coordination and utilization of resources still appears at the same value of z ($\equiv z_c$). Moreover, the scaling structure observed in non-evolutionary games also persists, although the precise form of the scaling function is different.

While the results of Ref. 6 are very interesting, there are clearly many situations in which restricting all agents to use strategies from the same strategy space, or restricting their evolutionary modifications to be drawn from the same strategy space is unrealistic. In this paper we extend our study of evolution in minority games to games in which different agents may have strategies drawn from different strategy spaces, and in which evolution may change the strategy space from which an agent’s strategies are drawn. In these games we find some surprising and robust features, notably that after evolution, the wealthiest agents are those whose strategies are drawn from the *smallest* strategy spaces (i.e. the smallest values of m). This is quite a counter-intuitive result. In addition, we find that for a given number of agents, N, the average agent wealth as a function of the size of the strategy space used by that agent (i.e., as a function of m), is roughly a step function. The step transition from wealthy to poor agents occurs at a value of m =

$m_t \approx m_c$-1, where $m_c$ is critical value of m at which the system achieves best emergent coordination for N agents in the games played with fixed m.

The outline of the rest of this paper is as follows: In the next section, we describe the evolutionary games studied in the paper, including a description of the various evolutionary algorithms used. In Section III we describe the most important results of our study. Section IV is devoted to some simple explanations of the results of Section III. The paper ends with Section V which contains a summary and discussion of our results.

## II. Evolutionary games with variable m

As in most previous work on minority games, we consider here a game with a fixed number of agents, N. At each time step of the game, each agent must join one of two groups (labeled 0 or 1). Each agent in the minority group at a given time step is rewarded with a point, each agent in the majority group gets nothing. In these games, the agents make their choice (to join group 0 or group 1) by following the prediction of a strategy. Strategies make their predictions by using information drawn from a set of common, publicly available information provided to all the agents at each time step. In the games studied here, that information is the list of which were the minority groups for the most recent past m time steps. Thus, a strategy is a look-up table with 2 columns and $2^m$ rows. The left-hand column contains a list of all $2^m$ possible common signals that the strategy can receive at a given time step of the game corresponding to the $2^m$ possible sequences of m 0's and 1's. For each such signal, the right-hand column contains a 0 or 1 which is that strategy's prediction of which will be the minority group in response to the given signal.

At the beginning of the game, each agent is randomly assigned s such strategies (in general, different, random sets of strategies for different agents). At each time step of the game, an agent must choose which of his s strategies to use. In the games studied in this paper, each agent keeps a running tally of how well each of his s strategies has done at predicting the correct minority group for all times since the beginning of the game. He then chooses to use that strategy that is currently doing the best. Ties among strategies may be broken in a variety of ways, the simplest being a random choice among the tied strategies.

Consider minority games in which different agents play with strategies that have different values of m. In the evolutionary versions of these games, agents with poorly performing strategies may also change the m-value of their strategies in response to selective pressure. For simplicity we consider the case in which all of an agent's s strategies have the same value of m at each time step of the game. (I.e., a given agent may not simultaneously use strategies of different m.) At the beginning of the game, strategies are distributed to the N agents so that $n_m(t=0)$ agents each have s (random) strategies of memory m. Clearly $\Sigma n_m(t)=N$. In all the games described here, we restrict ourselves to a universe in which $1 \leq m \leq 16$. We have studied games with various initial distributions of agents, $n_m(t=0)$. The main results we present below are independent of $n_m(t=0)$, although some details, such as rates of convergence do depend on $n_m(t=0)$. For specificity, unless stated otherwise, $n_m(t=0)$ is generally independent of m for the results reported below.

To complete the specification of the system, we must specify the evolutionary dynamics. There are many different ways to define evolutionary dynamics consistent with the notion of selective pressure. We have chosen to look at several which are associated with removal of poorly performing strategies. In this paper, we have not incorporated effects such as incremental mutation or reproduction. Studies including those dynamics will be reported elsewhere[11]. As we shall explain below, we find that some central features of our results are independent of the details of the evolutionary processes that we have studied. We believe that these features may be yet more general.

To evolve our system, we define a time, $\tau$, which is the duration of one generation. During $\tau$ time steps, the agents' strategies do not change. At the end of $\tau$ time steps, we rank the agents by wealth accumulated during that generation (i.e., how many times they have been in the minority group). We define a "poor" agent to be one whose wealth is in the lowest p percentile of agent wealth. We call p the "poverty level". We randomly choose half the agents whose wealth is in the lowest p percent of agents, and replace their s strategies with s new strategies. The new strategies do not necessarily have the same value of m as the strategies they replaced, so that $n_m$ is in general a function of the generation number. Those agents whose strategies are not replaced maintain the relative rankings of their strategies from one generation to the next. New strategies given to an agent at the beginning of a generation to replace poorly performing strategies start out with equal rankings. The game is played for an additional $\tau$ time steps, and the evolutionary process is repeated. In the results reported here, each agent has s=2 strategies, $\tau$ =20,000 time steps, and p is set so that the impoverished group is defined as either the poorest 10%, 20% or 40% of the population. We have studied two variants of these evolutionary dynamics. In the first, variant A, an agent chosen for strategy replacement is given strategies of any memory, m, with equal probability. In the second, variant B, an agent chosen for strategy replacement, whose strategies have memory m, is given strategies with memory m+1 or m-1, with equal probability.[12] Using these parameter ranges and the two evolutionary dynamics, we have studied a variety of games with N=101, 201, 401, and 801 agents run for a total of between 300 and 1800 generations (6 million to 36 million time steps). In all, we have performed about 140 experiments, the results of which are used in this paper.

## III. Results

In these games we will look at both the system-wide utilization of resources, and at the distribution of wealth to the agents. Since the strategies of different agents can have different values of m, it will be particularly interesting to see how wealth is distributed as a function of m.

### A. System-Wide Performance

We turn first to a description of the collective utilization of resources by the system. As in previous studies of minority games, it is convenient to consider, as an inverse measure of the goodness of resource utilization, $\sigma$, the standard deviation of the number of agents belonging to group 1. The smaller $\sigma$ is, the larger the typical minority is, and so the more points are awarded to the population *in toto*. In previous studies of minority games, we have found that $\sigma^2/N$ had very interesting scaling properties, and so it is the quantity we will consider here.

We have performed a variety of experiments with different values of N, p, and T, the total number of time steps in the game, using both variants of the evolutionary dynamics described above. In Figs. 1-3 we plot $\sigma^2/N$, averaged over the final 100 generations of the game, as a function of N for games played with different values of p. In each graph we have used different symbols to denote which of the two variants of evolutionary dynamics, A or B, were used. Note three major features of these figures:

1. $\sigma^2/N$ is approximately independent of N for each variant A and B.
2. For smaller p, $\sigma^2/N$ is lower for variant A than for variant B, but that difference disappears for p=40%.
3. $\sigma^2/N$ increases with increasing p.

As we shall discuss in the next section, observation 1 is a generalization of the scaling observed in the fixed m games[3,4,6]. Deviations from the N independence seen most strongly in Fig. 1 are most likely due to slow convergence and small values of T for some of the runs, as we shall discuss below. Observation 2 is most likely due to different rates of convergence for the two different variants. Indeed, lower values of $\sigma^2/N$ for variant B (especially for p=10%) are typically achieved for longer runs. Observation 3 mirrors a similar dependence on p in the case of evolutionary games with a fixed strategy space[5] for values of $m \neq m_c$.

It is also interesting to compare these values of $\sigma^2/N$ with those obtained in games with a fixed strategy space. In Fig. 4 we indicate some typical values of $\sigma^2/N$ for the variable strategy space games played here, and compare them with values for games played with a fixed strategy space. The results presented in this figure for the variable strategy space games are those obtained using variant A of the evolutionary dynamics, since convergence is faster, particularly for small p. In general, $\sigma^2/N$ is quite low for the variable strategy space games. $\sigma^2/N$ is much smaller than the non-evolutionary, fixed strategy space game, even for $m=m_c$. However, $\sigma^2/N$ for the variable strategy space games is not generally as small as that of the evolutionary, fixed strategy space games at $m=m_c$. For the latter, $\sigma^2/N$ is about 0.0245, (nearly) independent of p. When p is small (in our case p=10%), $\sigma^2/N$ is approximately equal to the value achieved in the evolutionary fixed m game at $m=m_c$,[6] but is larger for larger p.

B. Agent Wealth

To help understand the nature of evolution in these games, we consider how $n_m$ and $w_m$, the average wealth per agent accumulated in a given generation as a function of m, varies as the system evolves. In Fig. 5 we present a sequence of snapshots of one game illustrating the variation over time of both $n_m$ and $w_m$. In this example, N=401, the poverty level is set at 10% (so that about 20 agents change their strategies at the end of each generation), and the evolutionary dynamics is variant A. The initial distribution of agents among the m-bins, $n_m(t=0)$, is uniform (except for the last bin). This example was run for a total of 600 generations. The plots show $n_m$ and $w_m$ every 30 generations. The first snapshot, labeled generation 0 is $n_m$ and $w_m$ at the end of the first generation, before any evolution has taken place. In this and the next figure, $w_m$ is plotted in a normalized form to highlight the relative values of average wealth among agents in different m-bins. Thus, the maximum $w_m$ in a given generation is set to two and the minimum is set to one.

Note two important features late in evolution:

1. Average agent wealth as a function of m, $w_m$, is described by a near step function with high wealth accruing to agents with small m. In this example, the transition from large values of $w_m$ to small values of $w_m$ occurs at a value of m≈7.
2. Most agents have small values of m. $n_m$ also falls from relatively large to relatively small values at m≈7.

The step function behavior of $w_m$ is a very robust feature of evolution with a variable strategy space, and appears after a sufficiently long time in all our runs, regardless of the values of N, p, T, the variant of the evolutionary dynamics used, or the initial distribution of agents in m-bins ($n_m(t=0)$). The qualitative property that $n_m$ changes from large values for small m to small values for large m is also robust. However, the precise nature of the dependence of $n_m$ for small m depends on the nature of the evolutionary dynamics. For example, in fig. 6 we show another example of evolution in the same format as Fig. 5, but this time for N=401, p=10% and variant B of the dynamics In this figure, snapshots are shown every 90 generations, due to the slower convergence of this variant of the dynamics. Note that late in evolution $w_m$ is still a step function with a transition at a value of m≈6, $n_m$ is large for m≤6 and small for m>6, but the functional form of $n_m$ for m≤6 is different than in Fig. 5.

It is interesting to note that in generation 0 in both Figs. 5 and 6, $w_m$ peaks at about m=6. This is a typical feature of these kinds of runs, and is in marked contrast to the behavior of $w_m$ later after evolution. What is particularly striking about this is that agents in the low m bins initially do relatively poorly. Nevertheless, evolution (in combination with phase space arguments, as we shall explain below) ultimately selects solutions in which agents in the low m bins are wealthy.

The step function-like behavior of $w_m$ after evolution can be characterized by the value of m, $m_t$, at which the transition from wealthy to poor agents occurs. Since $w_m$ is monotonic, we define $m_t$ as a matter of principle to be that value of m at which $w_m$ is half its maximum value. However, since we have restricted ourselves to games with integer value of m, we can *operationally* define $m_t$ to be the first value of m, such that $w_m$ is less than half its maximum value. We denote this operationally determined value by $\tilde{m}_t$. In Fig. 7 we show a semi-log plot of $\tilde{m}_t$, averaged over several runs, as a function of N for games played with variant A of the dynamics. It is clear that there is a linear relationship between $\tilde{m}_t$ and log N. Since there is also a linear relationship between $m_c$ and log N in the fixed strategy space games, this suggests that there may be simple relationship between $m_t$ and $m_c$. We find that $m_t \approx m_c$-1, and $\tilde{m}_t \approx m_c - \frac{1}{2}$. Recall that for N=101 agents, $m_c \approx 5.2$, and increases by one everytime N is doubled. The results shown in Fig. 7 are thus consistent with the relation $\tilde{m}_t \approx m_c - \frac{1}{2}$. These relationships will be discussed further in the next section.

Another very robust feature of these games has to do with the wealth distribution of agents within m-bins. In Fig. 8 we show a typical scatter plot of agent wealth as a function of m for the $300^{th}$ generation of a game played with N=401,p=10%, and variant A of the evolutionary dynamics. We see that, although $w_m$ is roughly independent of m for $m<m_t$, the distribution of wealth within each m-bin broadens as $m \rightarrow m_t$ from below. This is a qualitatively robust finding and occurs in all of our simulations, independent of parameter settings and independent of the variant of the evolutionary algorithm.

## IV. Understanding the Results

The most robust features of minority games with variable strategy spaces are

1. After evolution, $\sigma^2$/N is generally independent of N
2. After evolution, $\sigma^2$/N is quite low, but, except for small values of p, is generally not as small as the value obtained for the evolutionary fixed strategy space game at $m=m_c$.
3. After evolution, agents tend to use strategies with small values of m, so that m-bins with $m<m_t$ are highly populated, but m-bins with $m>m_t$ are sparsely populated.
4. Average agent wealth in a given m-bin, $w_m$, is roughly a step-function, being high, and roughly independent of m for $m<m_t$, and low for $m>m_t$.

5. For a given number of agents, N, the transition from wealthy to poor agents occurs at a value $m_t \approx m_c$-1 (operationally, $\tilde{m}_t \approx m_c - \frac{1}{2}$), where $m_c$ is the critical value of m at which $\sigma^2$/N takes on its lowest value for the fixed strategy-space game played with N agents.
6. The spread in agent wealth within an m-bin increases as $m \rightarrow m_t$ from below.

At first sight these general results seem somewhat surprising for two reasons. First, given the fact that selection should work to improve individual and, possibly, system-wide performance, one might have expected that the system would evolve so that all agents would play with strategies with memory $m_c$. In addition, given that this is not the state to which the system evolves, it is furthermore surprising that the wealthiest agents are those with small memory ($m<m_t$).[13]

To understand this general structure, it is important to recognize a general principle which seems to be at work in these systems: The important quantity in determining the overall efficacy of the system is the ratio of the dimension of the strategy space available to the agents divided by the number of agents. When this ratio is about 1/3 the system does best at distributing resources. This is a very stable feature of minority games and is robust to many changes in the system including changes in the nature of the information set[14] and the introduction of evolution in games with fixed strategy spaces[6], or even the introduction of exogenous random driving signals.[9] In fact, the same principle is at work here. To a good approximation, after evolution, most agents are distributed in a global strategy space whose dimension is approximately $2^{m_c}$ which is the critical dimension of the strategy space for N agents. To see this, note that after evolution nearly all agents have values of $m \le m_t$. But the dimension of the strategy space for memory m is $2^m$ so the total *effective* strategy space available to the agents is (at least approximately) the direct product space of all the strategy spaces associated with memory $m \le m_t$, and so has dimension $\sum_{m=1}^{m_t} 2^m = 2^{m_t+1} - 2 = 2^{m_c} - 2 \approx 2^{m_c}$. Thus, evolution does indeed move the system toward a critical value of the effective available strategy space, but the qualitative nature of the space is much different than in the game restricted to a single value of m.[15] The observation that $\sigma^2/N$ is independent of N (Figs. 1-3) and that $m_t$ (and $\tilde{m}_t$) is proportional to log N (Fig. 7), is the analogue in the multi-m game of the scaling with z in the fixed m minority game: In the multi-m game, the system automatically picks out the value $m_t$ which plays the role of $m_c$ in these games. As for the difference between $m_t$ and $\tilde{m}_t$, we note that the operational definition states that $\tilde{m}_t$ is defined as the first m bin *after* $w_m$ has fallen to at least half its maximum value. This biases the definition of $m_t$ by +½ , thus producing the relationship between the operationally defined value $\tilde{m}_t$, and $m_c$. The theoretically interesting quantity is, nevertheless, $m_t$.

Given this, however, we can still ask why the system evolves toward this state, and not toward the state in which all agents sit in the strategy space of memory $m_c$? The reason is that there are many more states accessible to the evolutionary dynamics that have a distribution of agents in bins with $m<m_t$ than there are states with all agents in the single m-bin with $m=m_c$. This is fundamentally a phase space or entropic argument. Recall that under these evolutionary dynamics, an agent whose strategies are altered moves first to a different m-bin with some probability (which depends on the variant of the evolutionary dynamics), and then chooses strategies within this m-bin. Under such a scenario, it is clear that the probability of finding a large number of agents in the single m bin with $m=m_c$ is *a priori* much lower than the probability to find agents distributed in a range of m-bins. The minority dynamics is generally effective at distributing resources if the dimension of the effective strategy space is close to optimal, regardless of the nature of that strategy space. It is, therefore, much easier for the system to evolve to one of the many states in which agents are distributed in a variety of m-bins whose total dimension is $\sim 2^{m_c}$, rather than the single state in which all agents have memory $m_c$.[16] Of course, strict entropy also favors the (in general many) values of m for $m>m_c$. But since there are not enough agents to effectively coordinate choices, agents in those bins will not fare well, and occupancy there will be selected against by the evolutionary dynamics.

Although the evolutionary dynamics used here are clearly sensible and reasonable (and probably applicable to a wide variety of real systems in the social and biological sciences), there may be other evolutionary dynamics which favor evolution toward the singular state of occupancy at $m=m_c$. For example, if the agent's assignments into m-bins is weighted in some way (say, by either the dimension of the strategy space associated with that value of m, or by the number of strategies in that m-bin), then agents would be favorably placed in bins with large values of m. On the other hand, if they were placed in bins with too large a value of m, their performance would be poor, since they would not be able to coordinate their strategy choices well. Under such dynamics, it might be possible for the system to be driven to the state in which nearly all agents occupy the bin with $m=m_c$. The relevance of such dynamics will, of course, depend on the particular system being studied. But in any case, if the evolutionary dynamics is not artificially too constrained, we expect that the system will tend to a state in which the ratio of the dimension of the effective strategy space to the number of agents is about 1/3.

Although the ratio of the effective strategy space to the number of agents is close to optimal, overall performance of the system, as measured by $\sigma^2/N$ is, for most values of p, not as good as in the evolutionary fixed m case with $m=m_c$. The reason is that there is some continuing exploratory overhead. Since poorly performing agents will sometimes choose values of $m>m_t$, those agents will not be able to coordinate their choices very well, thus lowering system-wide performance. To see that this is the origin of the lowered performance, note first that the wealth of agents with $m>m_t$ is about what one would have for agents making random choices, thus indicating that, to a first approximation, those agents are choosing randomly between their two strategies. Since the wealthy agents and the poor agents fall into two distinct groups ($m<m_t$ and $m>m_t$, respectively), we can ask, what the system wide performance would be if we included only those agents with $m<m_t$ in the calculation of $\sigma^2/N$. Typically, we find that the value of $\sigma^2/N$ so computed is roughly consistent with the value of $\sigma^2/N$ at $z=z_c$ after evolution, for the fixed m case. Specifically, let $x_l$ be the number of agents in the minority group with $m<m_t$ and $x_g$ be the number of agents in the minority group with $m>m_t$. Also, let $N_g$ ($N_l$) be the number of agents with $m>m_t$, ($m<m_t$) and let $\xi = N_g/N$. Then, it is easy to show that

$$\sigma^2 = \langle(x_l + x_g - N/2)^2\rangle = \sigma_l^2 + \sigma_g^2 \qquad (4.1)$$

where

$$\sigma_l^2 = \langle(x_l - N_l/2)^2\rangle \text{ and } \sigma_g^2 = \langle(x_g - N_g/2)^2\rangle \qquad (4.2)$$

so that

$$\sigma^2/N = (1-\xi)\sigma_l^2/N_l + \xi\sigma_g^2/N_g. \qquad (4.3)$$

Because the agents with $m>m_t$ are unable to coordinate their choices, their average wealth is close to what one would expect for agents in the random choice game (RCG). It is reasonable, therefore, to use as the value for $\sigma_g^2/N_g$ 0.25, which is what we would find in the RCG. If we then use the observed value of $\sigma^2/N$ for the multi-m game, and the observed value for $\xi$, we can solve for $\sigma_l^2/N_l$. Doing this, we find that the computed value of $\sigma_l^2/N_l$ is consistent with the value of $\sigma^2/N$ at $z=z_c$ after evolution, for the fixed m case. Thus, the lowered performance of the system in the multi-m case is due to the evolutionary, exploratory overhead which at any time is expressed in the agents with $m>m_t$.

Another robust feature of our results is that fact, demonstrated in Fig. 8, that the spread in agent wealth increases as m approaches $m_t$ from below, even though $w_m$ stays roughly constant. The simple explanation for this is that as m increases the heterogeneity of the agents and their strategies also increases leading to a wider distribution in agent wealth. Of greater interest, of course, is the observation that $w_m$ is roughly independent of m for $m<m_t$. This just reflects the fact that all agents with $m<m_t$ are able to reasonably coordinate their minority choices since they dwell in an effective strategy space of roughly the right size (dimension $2^{m_c}$) for best utilization of resources.[17]

## V. Summary and Discussion

Summary

The main results of our investigation are that evolution in a mixed strategy space setting leads to states in which the most wealthy agents populate the lows m-bins. For a given total number of agents, N, the wealth per agent in an m-bin, $w_m$, is roughly a step function with a transition occurring at $m=m_t \approx m_c-1$, where $m_c$ is the critical value of m at which $\sigma^2/N$ is smallest in the fixed m game. This means that the effective size of the strategy space occupied by most of the agents is about $2^{m_c}$. Consequently, evolution in the minority game with variable strategy spaces for the agents typically leads to states which are critical in the sense that the ratio of the dimension of the effective strategy space to the number of agents playing the game is about 1/3, as in the fixed-m cases. In addition, after evolution, $\sigma^2/N$ is generally independent of N, and, once exploratory overhead is accounted for, $\sigma^2/N$ for the variable strategy space game has a value consistent with that found in the evolutionary fixed m game at $m=m_c$.

Discussion

The most remarkable features of our result are the twin robust findings that the agents that do the best are those with the lowest memory, and that, at the same time, the system evolves to a state that is critical in the sense that it is still characterized by the same effective critical size of the strategy space seen in the simpler games played with fixed m. The universality of the critical value of the strategy space is most impressive. But equally impressive is the fact that the system can manifest that criticality in a surprising way. After evolution, the agents and their strategies look nothing like the population in the fixed m game. In the multi-m case, wealthy agents are those with the simplest strategies, and the population is likewise distributed primarily in the low-m bins. On the other hand, agents with large m strategies do poorly. (We call this the “too clever by half” phenomenon.) In the fixed m case, at $m=m_c$ agents, perforce, all have the same memory and respond to exactly the same set of signals. Nevertheless, there is a deep commonality between these very different looking systems. In both cases the dimension of the effective strategy space is the same for a given number of agents and overall system performance is comparable.

We also see in our work consequences of evolutionary dynamics that are well known in other systems. First, there is the exploratory overhead which in our system manifests itself in the agents with memory $m>m_t$, and leads to some degradation of overall system performance. Second, the rate of convergence to a moderately stable macroscopic configuration varies, depending on values of some evolutionary parameters. In our case, convergence rates vary with the poverty level, p, as well as with the variant of evolutionary dynamics used. Also, the details of the state to which the system evolves may depend to some extent on the initial conditions and on the evolutionary dynamics. For example, in the experiments discussed here, if the initial population of agents is concentrated in one m-bin with $m>m_t$, and if the evolutionary dynamics allows only changes of ±1 in an agent's memory with each generation, then there appears to be a tendency for $n_m$ to increase as m approaches $m_t$ from below. (It is, of course, possible that this is also a transient effect, but we have seen such a feature persist in runs of up to 900 generations.)

Our results also raise deep questions about the interaction between the fundamental principles that govern evolution and the characteristics of the space of possible outcomes. Our work here, coupled with previous work on the minority game, suggests that there may be an important and fundamental principle at work in the evolution of systems in which adaptive agents compete for scarce resources. Namely, while the rules of the game are set up to ensure that agents seek to optimize their own utility, evolution also pushes the system to a configuration in which the global good is optimized (or nearly so). We emphasize that this seems to be a non-trivial consequence of evolution in the systems we have studied here, and we speculate that it is a very general and important priniciple of evolutionary adaptive competition. On the other hand, the specific way in which the underlying priniciple is manifested in a specific situation may depend on a various parameters governing the dynamics of that situation. For example, in the minority game we have studied, one important feature of the underlying strategy space is that there are many more ways to distribute agents over several low m bins than to distribute them all in one $m=m_c$ bin. This is a major reason why systems described in this paper evolve toward states with highly populated low m bins. More generally, the nature of the phase space is likely to be one important factor channeling the dynamics of all evolutionary systems. Other factors that are likely to be important in determining the precise nature of the evolved state include the payoff structure[18], developmental and historical constraints[19] and various dynamical relationships among agent types[20].

Our study of evolution in minority games has helped highlight what we believe to be a general priniciple that may play an important role in evolving social and biological systems. But there is much yet to be done in order to understand more deeply the generality and nature of this principle, as well as the ways in which it can be manifest and its limitations.

---

## Footnotes

[1] W. Brian Arthur, *Amer. Econ. Assoc. Papers and Proc.* **84**, 406 (1994).

[2] D. Challet and Y.-C. Zhang, *Physica A*, **246**, 407 (1997).

[3]R. Savit, R. Manuca andR. Riolo, Phys. Rev. Lett. **82**, 2203 (1999).

[4] R. Manuca, Y. Li, R. Riolo and R. Savit, *The Structure of Adaptive Competition in Minority Games*., UM Program for Study of Complex Systems Technical Report PSCS-98-11-001, at http://www.pscs.umich.edu/RESEARCH/pscs-tr.html, or LANL eprint at http://xxx.lanl.gov/abs/adap-org/9811005, submitted for publication.

[5] D. Challet and Y. Zhang, Physica A **256**, 514 (1998).

[6]Y. Li, R. Riolo and R. Savit, *Evolution in Minority Games I: Games with a Fixed Strategy Space*, UM Program for Study of Complex Systems Technical Report PSCS-99-03-001, at http://www.pscs.umich.edu/RESEARCH/pscs-tr.html, or LANL eprint at http://xxx.lanl.gov/abs/adap-org/9903008, submitted for publication.

[7] N. F. Johnson, M. Hart, and P. M. Hui, *Crowd effects and volatility in a competitive market*, LANL eprint at http://xxx.lanl.gov/abs/cond-mat/9811227.

[8] M. A. R. de Cara, O. Pla, and F. Guinea, *Competition, efficiency and collective behavior in the "El Farol" bar model*, LANL eprint at http://xxx.lanl.gov/abs/cond-mat/9811162.

[9] A. Cavagna, *Irrelevance of memory in the minority game*, LANL eprint at http://xxx.lanl.gov/abs/cond-mat/9812215.

[10] A. Cavagna, J. Garrahan, I. Giardina and D. Sherrington, LANL eprint at http://xxx.lanl.gov/abs/cond-mat/9903415 (1999); D. Challet, M. Marsili and R. Zecchina, LANL eprint at http://xxx.lanl.gov/abs/cond-mat/9904392 (1999).

[11] A. Sullivan, Y. Li, R. Riolo and R. Savit, in preparation.

[12] Boundaries are treated in the following way: If m=1(16), 50% of the agents chosen for replacement are given strategies with memory 2(15) and the remaining 50% are given strategies with memory 1(16).

[13] In this regard, our conclusions differ from those of Challet and Zhang (Refs. 2 and 5) , who assert that there is always an evolutionary advantage to having more memory in a multi-m game, and that agents with larger memory always out-perform agents with smaller memory. Although we believe the results of their simulations are correct, it appears that the inference drawn from them is too general. As we have seen, agents who live in a strategy space which is too large to coordinate, given the number of players in the game, do substantially worse than players with smaller memory.

---

[14] Y. Li, R. Riolo and R. Savit, in preparation.

[15] In this regard, it is also worth noting that agent wealth is strongly correlated with an agent's average distance in behavior space (see Ref. 4 for a definition of behavior space) from all other agents (most of whom have $m<m_t$), but is not correlated with an agent's distance in behavior space from only the other agents with the same value of m. This suggests that it is more appropriate to think about all agents living in the space which is the direct product of the strategy spaces with $m<m_t$, rather than thinking about a collection of agents with different m's playing different games and competing only with other agents having access to the same information.

[16] Here we also note that the precise distribution of agents in m-bins $\{n_m\}$ does not appear to be universal, again pointing to the fact that many different states may be occupied all of which share the same general profile of wealth as a function of m.

[17] There is, in general, a small decrease in $w_m$ as m increases toward $m_t$. However, this variation is much smaller than the difference between $w_m$ for m below and above $m_t$.

[18] For an example of a relatively recent discussion of this issue, see S.Kaufmann, *The Origins of Order: Self-Organization and Selection in Evolution*, (Oxford Univ. Press, NY, 1993).

[19] See, for example, L. W. Buss, *The Evolution of Individuality*, (Princeton Univeristy Press, 1987); W. Fontana and L. W. Buss, SFI Working Paper 93-10-067 (1993).

[20] W. Fontana, Functional Self-Organization in Complex Systems, in *1990 Lectures in Complex Systems, SFI Studeis in the Science of Complexity, Lecutre Notes, Vol. III*, L. Nadel and D. Stein (eds.), p. 407 (Addison-Wesley, 1991).

## Figure Captions

Fig. 1. $\sigma^2/N$ as a function of N for p=10%. Crosses indicate games played with variant A of the evolutionary dynamics, and circles indicate results for games played with variant B.

Fig. 2. $\sigma^2/N$ as a function of N for p=20%. Crosses indicate games played with variant A of the evolutionary dynamics, and circles indicate results for games played with variant B.

Fig. 3. $\sigma^2/N$ as a function of N for p=40%. Crosses indicate games played with variant A of the evolutionary dynamics, and circles indicate results for games played with variant B.

Fig. 4. A comparison of values of $\sigma^2/N$ for games with variable and fixed strategy spaces. In this graph results are plotted for games played wih N=101. The results plotted for the variable strategy space games are for variant A of the evolutionary dynamics. Similar values are obtained for other values of N, and indicated in Figs. 1-3.

Fig. 5 $w_m$ and $n_m$ every 30 generations for a game played with N=401 agents, p=10% and variant A of the evolutionary dynamics. The initial conditions are that $n_m(t=0)$ is uniform for $1 \leq m \leq 16$ (with the exception of bin 16), which comprises the universe of this simulation. Values for $w_m$ are scaled so that the maximum value in any generation is set to two and the minimum value is set to one. Generation 0 shows the results after the first generation of the run, before any evolution.

Fig. 6 $w_m$ and $n_m$ every 90 generations for a game played with N=401 agents, p=10% and variant B of the evolutionary dynamics. The initial conditions are that $n_m(t=0)$ is uniform for $1 \leq m \leq 16$ (with the exception of bin 16), which comprises the universe of this simulation. Values for $w_m$ are scaled so that the maximum value in any generation is set to two and the minimum value is set to one. Generation 0 shows the results after the first generation of the run, before any evolution..

Fig. 7. $\tilde{m}_t$, the operationally defined version of $m_t$, versus N on a semi-log plot. These results are for variant A of the dynamics. The values shown represent averages over a total of 24 runs (4 each for N=101 and N=801, and 8 each for N=201 and N=401)

ranging in duration from 6 million to 18 million time steps (300 to 900 generations). For 12 of the runs p=20%, and for 12 of the runs p=40%.

Fig. 8. Typical scatter plot of agent wealth as a function of m. Shown is a scatter plot for the 300th generation of a run in which N=401, p=10%, and the evolutionary dynamics is variant A. Note that in the low m bins there are many more agents clustered near w=10,000 than there are outliers, or than there are agents in the high m bins. This is not necessarily apparent on this figure since many black dots representing different agents are superimposed on each other near w=10,000.